# Tailoring Light-Matter Interaction with a Nanoscale Plasmon Resonator

Nathalie P. de Leon,[1,2] Brendan J. Shields,[2] Chun L. Yu,[1] Dirk Englund,[2] Alexey V. Akimov,[2,3] Mikhail D. Lukin,[2]* and Hongkun Park[1,2]*

[1]Department of Chemistry and Chemical Biology, and [2]Department of Physics, Harvard University, Cambridge, MA 02138, [3]Lebedev Institute of Physics, Moscow, Russia

*To whom correspondence should be addressed. Email: hongkun_park@harvard.edu, lukin@physics.harvard.edu

**We propose and demonstrate a new approach for achieving strong light-matter interactions with quantum emitters. Our approach makes use of a plasmon resonator composed of defect-free, highly crystalline silver nanowires surrounded by patterned dielectric distributed Bragg reflectors (DBRs). These resonators have an effective mode volume ($V_{eff}$) two orders of magnitude below the diffraction limit and quality factor ($Q$) approaching 100, enabling enhancement of spontaneous emission rates by a factor exceeding 75 at the cavity resonance. We also show that these resonators can be used to convert a broadband quantum emitter to a narrowband single-photon source with color-selective emission enhancement.**

Techniques for controlling light-matter interactions in engineered electromagnetic environments are now being actively explored. Understanding these interactions is not only of fundamental importance but also of interest for applications ranging from optical sensing and metrology to information processing, communication, and quantum science [1-3]. To enhance the coupling between an optical emitter and a desired mode of the radiation field, two approaches can be used [4]. One strategy is to increase the lifetime of the confined optical excitation in high-Q dielectric resonators, such as whispering gallery structures, micropillars, and photonic crystals [3, 5]. Another strategy is to reduce the effective mode volume ($V_{eff}$) of confined radiation [6-9], as is currently explored using plasmonic nanostructures capable of confining light to dimensions well below the diffraction limit.

    Plasmonic resonators, which combine the benefits of both strategies, have potential for engineering light-matter interaction at nanoscales and achieving large coupling between an emitter and the radiation field [2, 10]. However, experimental realization of these structures has remained an outstanding challenge [2, 11, 12]. Prior attempts to control and engineer surface plasmon polariton (SPP) propagation have relied on patterning of metal films using techniques such as focused ion beam and reactive ion etching [13, 14]. Unfortunately, these patterning methods introduce defects that act as scattering centers. Moreover, standard metal deposition techniques typically generate polycrystalline films with short SPP propagation lengths. Consequently, plasmonic cavities demonstrated thus far typically had large mode volumes to minimize absorption and scattering in the metal [15-17].

In this Letter, we propose and experimentally demonstrate the new strategy to realize a plasmonic resonator with an exceptionally small mode volume and a moderate Q that can drastically modify the interaction between a quantum emitter and SPPs. Our new approach, illustrated in Figure 1, takes advantage of chemically synthesized silver nanowires[18] for tight confinement and reduced group velocity of optical radiation. Owing to their high crystallinity, these NWs support propagation of surface plasmon polaritons (SPPs) over several micrometers in the visible range [16]. We define plasmon distributed Bragg reflectors (DBRs) and cavities by patterning polymethylmethacrylate (PMMA), a low-index dielectric. Unlike metal/air structures that have been previously employed to manipulate SPP propagation [13-17], patterned PMMA does not suffer from high scattering losses because the relatively low index contrast (compared to metal/air boundaries) reduces susceptibility to lithographic imperfections.

Thin silver NWs surrounded by air support a fundamental SPP mode whose spatial extent is on the order of the wire radius [8]. The effective refractive index for this mode increases when the surrounding medium changes from air to PMMA ($n_{eff}^{PMMA} > n_{eff}^{air} > 1$), providing the index contrast needed to define DBRs for SPPs. FDTD simulations show that a quarter-wave stack composed of several PMMA slabs reflects the incoming SPPs with > 90% efficiency (Fig. 1(c)). The stopband (the wavelength range over which the quarter-wave stack acts as an efficient plasmon mirror) can be tuned over the entire visible range by varying the stack period, with a bandwidth exceeding 100 nm. A sharp resonant feature appears within the stopband when two DBRs are placed together to define a plasmon cavity. Specifically, for a 100-nm diameter NW with DBRs composed of 6 PMMA slabs, a plasmon cavity with $V_{eff} \sim 0.04\ (\lambda/n)^3$ and $Q \sim 100$ can be achieved (Fig. 1(d), see also Supplementary Material).

When a quantum emitter is placed within this plasmon cavity, its spontaneous emission rate can be dramatically modified. The Purcell factor ($F$), which is defined as the ratio of the spontaneous emission rates within the cavity ($\Gamma$) and in free space ($\Gamma_0$), scales as $Q/V_{eff}$ (see Supplemental Material). Equivalently, $F$ can also be expressed as $F_0\mathfrak{I}$, where $F_0$ and $\mathfrak{I}$ are the Purcell factor of a bare silver NW and the cavity contribution, respectively. We note that $\mathfrak{I}$ is also the finesse of the cavity. For a plasmon cavity with a 100-nm diameter silver NW (Fig. 1(d)), $F_0$ ranges from 1 to 10 depending on emitter placement [8], and $\mathfrak{I} \sim 20$. $F$ can therefore be as high as 200 when a quantum emitter is placed at the peak electric field of the cavity mode.

Our plasmon resonator architecture offers distinct advantages over other photonic and plasmonic cavity structures [3, 15, 19, 20]. First, the Purcell enhancement achievable in our architecture improves dramatically as the device dimension is pushed well below the diffraction limit [8]. As the NW radius ($R$) decreases, $V_{eff} \propto R^3$, while $Q$ can be kept constant by choosing the cavity length to be the half the SPP wavelength (see Supplemental Material). Therefore, the Purcell enhancement also scales as $1/R^3$, indicating that extraordinarily strong coupling can be achieved for small diameter NW devices. This is in stark contrast to dielectric photonic waveguides, in which the field confinement decreases exponentially when the structure dimensions shrink below the diffraction limit [21, 22].

In addition, due to its ultrasmall mode volume, our resonator can be used to direct the emission of a broadband quantum emitter into a single cavity mode whose resonant wavelength is selected by the cavity design. When a broadband emitter (e.g. a solid state emitter with a broad phonon sideband) is coupled to a cavity with a much narrower resonance, the total Purcell factor becomes independent of $Q$ and increases only when $V_{eff}$ is decreased (see Supplemental Material). Therefore, in this broadband emitter regime, ultra-small mode volume plasmon resonators provide the only means to achieve efficient single photon sources in which the color of emission can be selected, and the rate into that mode can be enhanced over emission into free space.

We realize the plasmon resonator experimentally by first spin-coating PMMA on a Si/SiO$_2$ wafer, drop-casting silver NWs, and then spin-coating another layer of PMMA. Electron beam lithography and subsequent development yield suspended NWs in periodic PMMA slabs (Fig. 1(a)) [19, 23]. These plasmon DBRs and cavities were characterized via transmission measurements of single nanostructures as a function of wavelength. Due to the wavevector mismatch between propagating SPPs and free space photons, SPPs couple to the far field only at defects or wire ends. We obtain a transmission spectrum by focusing a supercontinuum laser to a diffraction-limited spot at one end of the NW, and recording the scattered intensity at the other end as a function of wavelength using a charge-coupled device (CCD) (Fig. 2(a)).

The data in Fig. 2(b) clearly show that the plasmon DBR exhibits a stopband in transmission, as predicted by the simulations. By comparing the transmission intensity just inside and outside of the stopband (over which range the propagation losses, collection efficiency, and in/out-coupling efficiencies should be constant, see Supplemental Material), we estimate the reflectivity of the DBR to be 90-95%. The drop-offs in intensity at shorter and longer wavelengths are due to material absorption and the near-IR cutoff of our optics, respectively.

Transmission spectra of cavities show a peak in the middle of the stopband. In the device shown in Fig. 2(c), the peak is at 638 nm and the full width at half maximum is 11 nm, corresponding to a $Q$ of 58. The highest $Q$ observed to date in our plasmon cavities is 94, close to the theoretically simulated maximum value. The transmission intensity on resonance is attenuated compared to that outside of the stopband due to higher absorption losses caused by the longer effective path length on resonance, $l_{eff}$. From the measured value of $Q$, we determine that $l_{eff}$ is ~5 μm (see Supplemental Material). This value is comparable to the SPP propagation length in bare silver NWs [24], and indicates that losses in our resonators are dominated by material absorption.

We next demonstrate the utility of our cavities to control both the color and rate of spontaneous emission of solid-state optical emitters. First, we show emission modification in an ensemble of CdSe quantum dots. These quantum dots were coupled to the devices by mixing them homogeneously in PMMA before fabrication. A fluorescence spectrum of the quantum dots coupled to the NW was obtained by exciting one NW end with green light ($\lambda$ = 532 nm) and collecting quantum dot fluorescence at the other NW end. Fig. 3(a) clearly shows that the plasmon-coupled emission is narrowed and shifted toward the resonance peak when compared to the fluorescence spectrum of uncoupled quantum dots.

The fluorescence lifetime is also modified by the plasmon cavity. The emission from uncoupled quantum dots exhibits a single-exponential decay, characterized by a lifetime ($\tau_{free}$) of 16 ns ± 3 ns. In contrast, the emission from quantum dots coupled to the NW exhibits a multiexponential decay because the quantum dots are distributed throughout the PMMA, and the detected emission originates from an ensemble of quantum dots along the NW. Notably, the initial slope of this decay yields the shortest lifetime ($\tau_{coupled}$) of less than 250 ps (Fig. 3(b)). This value suggests that the largest effective Purcell factor $F_{eff} = \tau_{free}/\tau_{coupled}$ is >75, despite the detuning of the cavity resonance relative to the peak of quantum dot emission. It is difficult to disentangle effects of nonradiative decay from lifetime data alone, but we note that quantum dots coupled to a silver NW in unpatterned PMMA show a multiexponential decay with initial slope corresponds to a lifetime of 4 ns. This evidence indicates that the much shorter lifetime observed in the cavity-coupled decay is due to radiative emission enhancement rather than nonradiative decay (see Supplemental Material).

We next demonstrate control over emission properties of individual diamond nitrogen vacancy (NV) centers using plasmon resonators. Diamond nanocrystals were coupled to silver NWs by co-depositing them during fabrication. Approximately 10% of nanocrystals exhibited stable, broad fluorescence characteristic of NV centers (Fig. 4(b)) [25]. Once a single NV coupled to the wire was identified, the resonator structure was defined by electron beam lithography around the NV center.

Fig. 4(a) shows scanning confocal microscope images of a resonator device with a coupled NV. The top panel shows a reflection image of the device, and the middle panel shows the fluorescence image recorded as the laser is scanned over the device. When the NV is excited (circled), using an independently scanning collection channel (bottom panel) we observe fluorescence from three locations: one corresponding to direct emission from the NV, and two corresponding to the ends of the NW. The direct emission from the NV exhibits strongly anti-bunched autocorrelation (inset of Fig. 4(b)), indicating that it originates from a single NV center. Photon-correlation measurements between the NV fluorescence spot and the NW end also show strong anti-bunching (inset of Fig. 4(c)), showing that the emission from the wire end also originates from the NV [6, 9]. Assuming an SPP out-coupling efficiency of ~5% for these NWs [6, 24], we estimate that 50-60% of the emission couples into SPPs for a typical device.

The NV fluorescence changes drastically when it is coupled to the plasmon cavity. Before resonator fabrication, the plasmon-coupled NV fluorescence exhibits a broad spectrum spanning a range of 630-740 nm (small, superimposed Fabry-Perot oscillations originate from scattering by the NW ends: see Supplemental Material) [9, 16]. After resonator fabrication, the plasmon-coupled NV spectrum exhibits a peak on resonance with the cavity mode and suppressed fluorescence within the stopband (Fig. 4(c)). The peak position can be placed anywhere across the NV fluorescence spectral range by changing the cavity design, thereby enabling the selection of the wavelength of single photon emission. The plasmon-coupled fluorescence intensity outside of the stopband ($\lambda > 720$ nm), $I_{out}$, is essentially unaltered by the cavity, and gives the baseline SPP-coupled fluorescence. Notably, the fluorescence intensity on resonance ($\lambda = 637$ nm), $I_{res}$, is higher than $I_{out}$. In contrast, the transmission spectrum measured by launching SPPs at the wire end with the supercontinuum laser shows that the transmitted intensity on resonance, $T_{res}$, is lower than that outside of the stopband, $T_{out}$.

Comparison of these intensities gives the radiative Purcell enhancement due to the cavity on resonance $\Im = (I_{res}/I_{out})(T_{out}/T_{res})$; for the device shown, we find that $\Im$ is $11 \pm 3$. This resonant enhancement is in addition to a broadband enhancement, $F_0$, due to the bare silver NW, which is estimated to be in a range of 1.5-2.5 for NWs of these dimensions [6, 9]. Combining these factors together ($F = F_0\Im$), we estimate the overall Purcell enhancement to be as high as 35 at the resonance peak, a value that exceeds the largest Purcell enhancement reported to date for NV centers coupled to dielectric cavities [26, 27]. We note that the observed Purcell enhancement is still lower than the theoretical maximum value expected for these devices, most likely because the NV center is not located optimally within the cavity.

The strong emitter-cavity coupling observed in the present study can be improved in several ways. Precise placement of NV centers at the peak electric field of the cavity mode would ensure maximum Purcell enhancement for a given device. Furthermore, because the Purcell enhancement scales as $1/R^3$, it can be made substantially higher using thinner wires [7, 8]. In the present study, we were limited to larger (~100 nm) diameter NWs because we relied on far-field excitation and detection of SPPs. While out-coupling to the far field is less efficient in thinner wires, efficient coupling to thinner NWs can be accomplished with near-field techniques such as evanescently-coupled optical fibers [28] and electrical detection [24, 29]. Other resonator geometries, such as those that make use of recently developed hyperbolic metamaterials can potentially be used to further enhance the coupling [30].

The realization of nanoscale plasmon resonators with exceptionally small mode volumes and high quality factors opens new possibilities for integrated plasmonic systems, novel realization of nanoscale lasers and spasers [31], sub-diffraction sensing, and optical interfacing of solid-state qubits. For instance, color-selective single photon sources could have applications in quantum cryptography, and these resonators can be used to direct NV emission into the zero-phonon line for coherent optical manipulation, a crucial requirement for the realization of such applications as single photon transistors [32]. Other possibilities include high spatial resolution imaging and enhanced coupling of individual molecules. Furthermore, the use of patterned, low-loss dielectrics for controlling SPP propagation in nanoscale plasmonic structures can potentially be extended towards other applications such as plasmonic circuit elements [33, 34], out-coupling gratings [11], and meta-materials [35, 36].

We acknowledge J. T. Robinson, A. L. Falk, F. Koppens, and J. D. Thompson for helpful discussions. We also gratefully acknowledge support from NSF, DARPA, the Packard Foundation, and the NSF and NDSEG graduate research fellowships (NdL).

**Figure Captions**

FIG. 1 (color online). (a) Schematic of device concept (top) and SEM image using an Inlens detector (bottom) of the DBR resonator fabricated around a silver NW. The sample is tilted at 30° from two different orientations to demonstrate that the NW is suspended from the substrate in PMMA. Scale bar = 1 µm. (b) Simulated electric field intensity of a plasmonic cavity at wavelengths inside the stopband (top left), outside the stopband (bottom left) and on resonance (right). (c) Simulated transmission spectrum of a DBR consisting of a 100 nm silver NW and PMMA slabs with a period of 200 nm. A quarter-wave stack composed of 6 PMMA slabs reflects the incoming SPPs with a reflectivity exceeding 90%. (d) Simulated transmission spectrum of plasmon cavity composed of two DBRs. The nominal cavity length corresponds to $\lambda_{SPP}$, and the linewidth gives a $Q$ of 100.

FIG. 2 (color online). (a) CCD image of SPP propagation overlaid on an SEM image of a device. (b) Transmission spectrum of plasmonic DBR. The minimum intensity in the stopband indicates a reflectivity of 90-95%, and sideband oscillations can be seen outside the stopband. (c) Transmission spectrum of plasmon resonator. The peak in the stopband at 638 nm has a width of 11 nm, corresponding to a $Q$ of 58. SEM images of both devices are shown in the insets, with scale bars that correspond to 1 µm.

FIG. 3 (color online). (a) Fluorescence spectrum of CdSe quantum dots in the substrate (blue, dashed) and coupled to the plasmon resonator (red). The transmission spectrum of the device is overlaid (gray, crosses). The fluorescence of coupled quantum dots is shifted and narrowed by the cavity resonance. (b) Lifetime measurements obtained by excitation with a pulsed laser of coupled (red) and uncoupled (blue, dashed) quantum dots. Uncoupled quantum dots have a lifetime of 16 ns, while coupled quantum dots show multi-exponential behavior, with the initial slope corresponding to a lifetime of less than 250 ps.

FIG. 4 (color online). (a) Scanning confocal microscope images of the device. The top panel shows an image of the reflected green laser light. The middle panel shows fluorescence in the red as the laser is scanned over the area. When the laser is focused onto the NV in the cavity (indicated with a white circle), an independent collection channel is scanned over the area to detect fluorescence (bottom). The bright spot in the center results from direct excitation and detection of the NV fluorescence. Two additional spots can be seen from the ends of the wire, where SPPs excited by the NV scatter into free space. Scale bar = 1 μm. (b) The fluorescence spectrum from the center of the NW shows broadband emission characteristic of NVs. The inset shows autocorrelation of the fluorescence, which is strongly anti-bunched, indicating that emission results from a single NV center. (c) The fluorescence spectrum from the end of the NW (red) shows significant modification, which corresponds to the transmission spectrum of the device (black). Cross-correlation between the SPP-coupled emission and the emission collected directly from the NV indicates that the fluorescence at the end of the NW originates from the same NV center (inset).

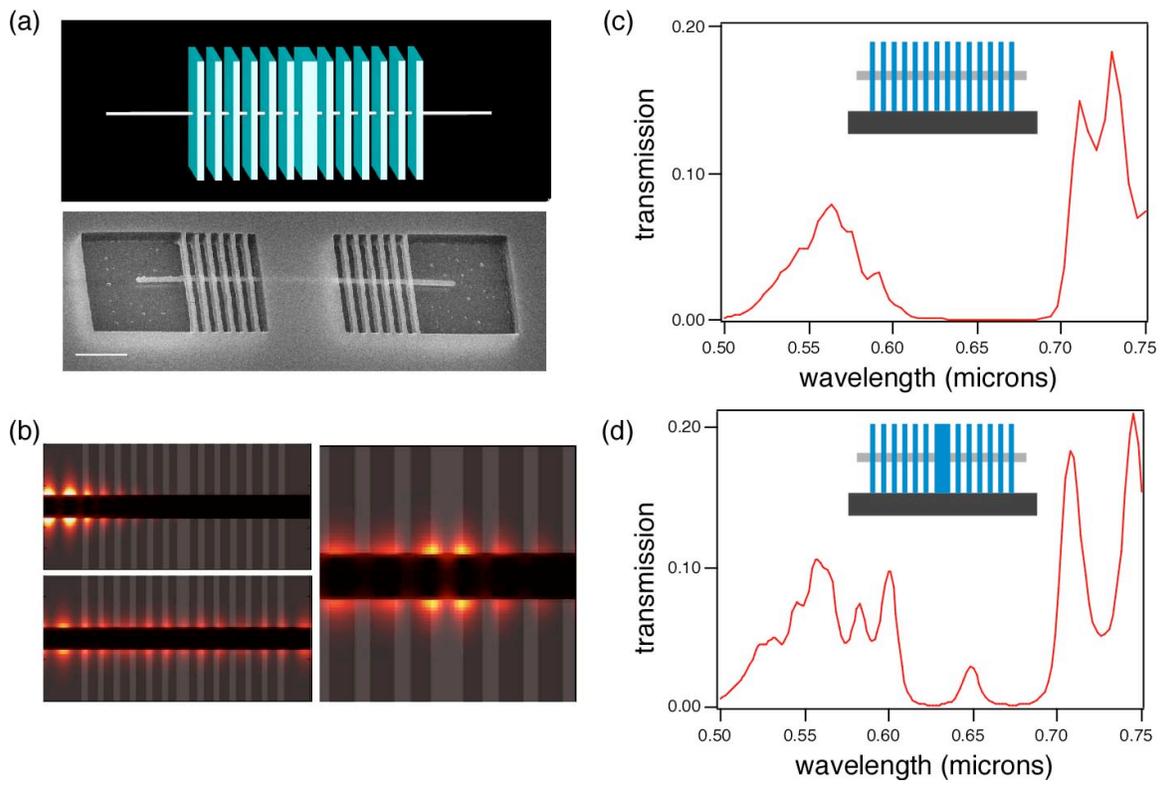

Figure 1

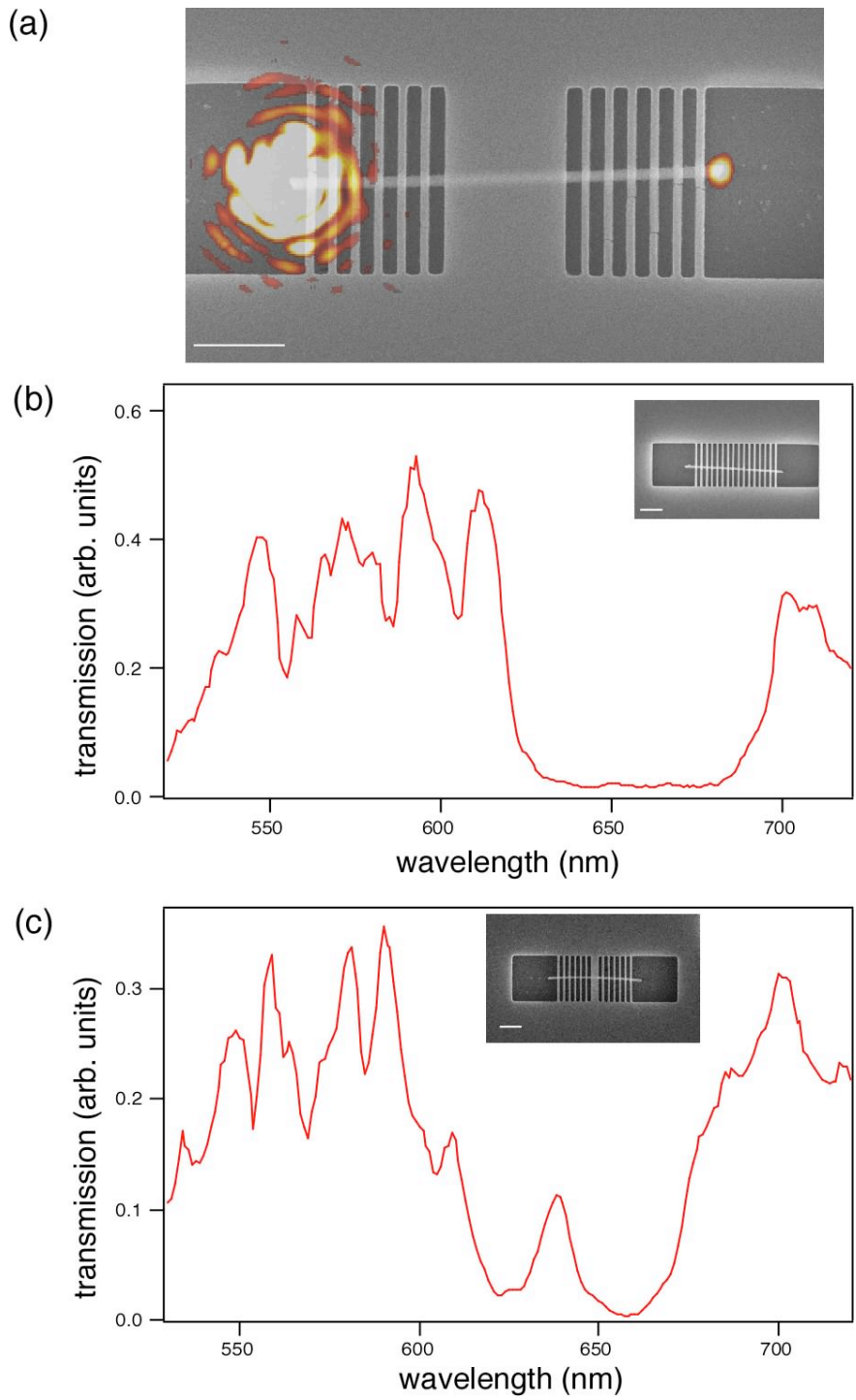

Figure 2

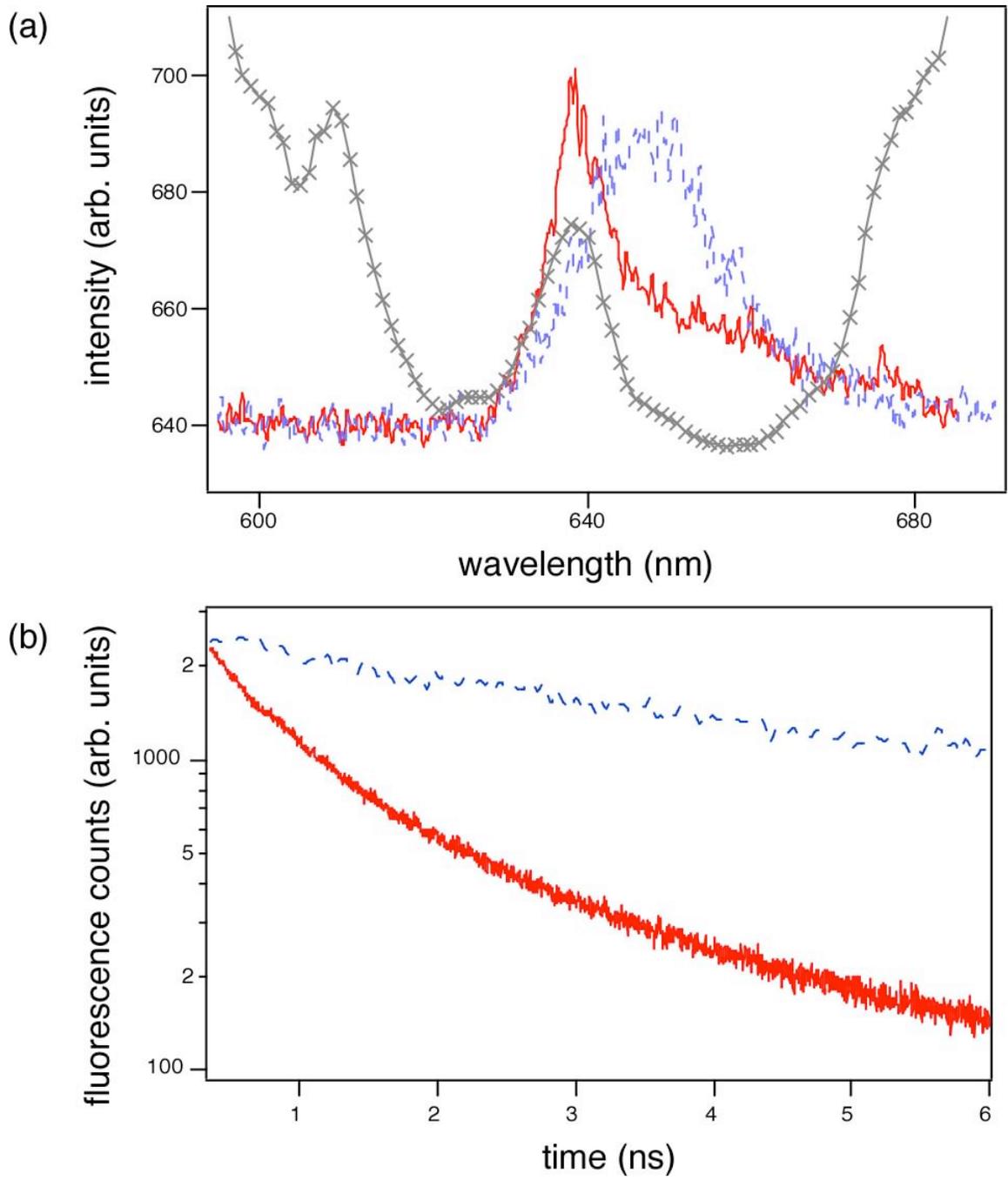

Figure 3

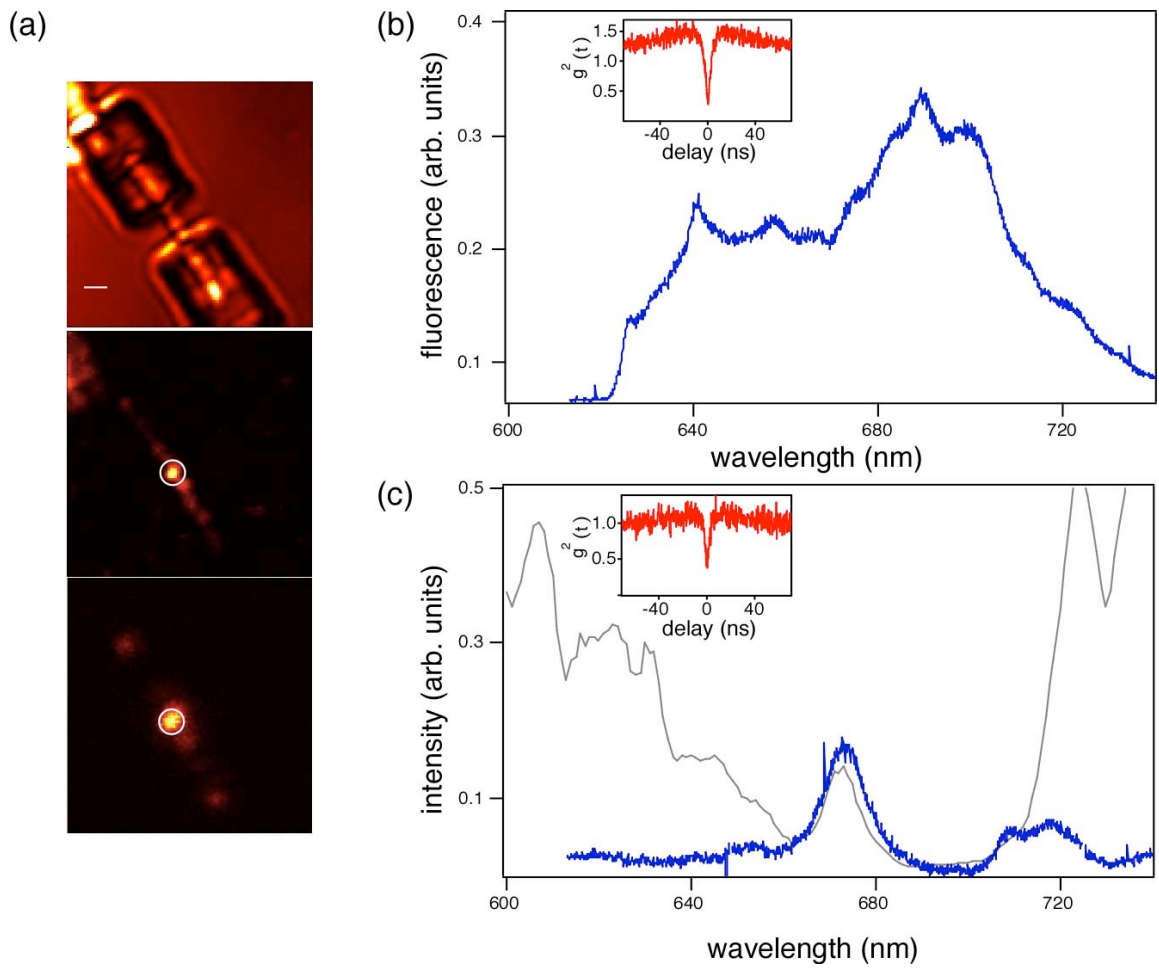

Figure 4

**Supplementary Material for Tailoring Light-Matter Interaction with a Nanoscale Plasmon Resonator**

Nathalie P. de Leon[1,2], Brendan J. Shields[2], Chun L. Yu[1], Dirk Englund[2], Alexey V. Akimov,[2,3] Mikhail D. Lukin[2], and Hongkun Park[1,2]

[1]Department of Chemistry and Chemical Biology, and [2]Department of Physics, Harvard University, Cambridge, MA 02138, [3]Lebedev Institute of Physics, Moscow, Russia

**A. Materials and Methods**

The silver nanowires used in our experiments were synthesized using a modified solution-phase polyol method described previously [1]. Resonators were fabricated by spin-coating 950 PMMA C4 (MicroChem) on a Si substrate with 300 nm $SiO_2$, dropcasting silver nanowires in ethanol, then spin-coating another layer of PMMA. Once silver nanowires were identified, the dielectric stacks were defined by e-beam lithography (Raith) and developed in methylisobutylketone and isopropanol.

FDTD simulations were performed using commercial software (Lumerical) with literature values for material constants [2].

Transmission spectra were taken using a Koheras SuperK supercontinuum laser coupled to an acousto-optic tunable filter, enabling the excitation wavelength to be selected. The beam was spatially filtered using a single mode fiber (NKT) and then focused to a diffraction-limited spot using an objective (100x, 0.8 numerical aperture).

Fluorescence spectra were taken using a CCD and a spectrometer (Princeton Instruments).

CdSe quantum dots (Invitrogen, 650 nm emission wavelength) with organic capping ligands were flocculated out of decane and dispersed in PMMA. This solution was then used for fabrication of resonators. Diamond nanocrystals (Microdiamant, 0-0.05 micrometres) were centrifuged, washed, and dispersed in ethanol before co-depositing with silver nanowires. We tuned the relative concentrations of nanowires and diamond nanocrystals so that approximately 30% of the wires had a single NV coupled to them, thus minimizing the probability that multiple NVs were coupled to one wire.

Lifetime measurements were performed using a frequency-doubled picosecond 1060 nm laser (Fianium, 400 fs pulsewidth, 80 MHz repetition rate down-sampled to 20 MHz) an avalanche photodiode with 50 ps timing resolution (Picoquant) and a fast photodiode with 80 ps timing resolution (Newport). Photon correlation measurements were taken using avalanche photodiodes (Perkin Elmer). The details of the experimental setup used for photon correlation measurements have been described previously [3].

**B. Plasmonic DBR characterization**

In our supercontinuum laser transmission measurements, the observed intensity ($I$) is given by

$$I(\lambda) = I_0(\lambda) T(\lambda) \exp\left(\frac{-l}{l_0(\lambda)}\right) \eta_c(\lambda) \eta_{pl}(\lambda), \tag{1}$$

where $I_0$ is the incident intensity, $T$ is the transmission through the DBR, $l$ is the optical path length along the NW, $l_0$ is the characteristic SPP propagation length in the NW waveguide, $\eta_c$ is the collection efficiency, and $\eta_{pl}$ is the product of the in-coupling and out-coupling efficiencies of the NW.

We fabricated a variety of plasmonic DBR devices with different periods and nanowire diameters in order to tune the stopband across the visible range. The stopband position and device dimensions can be used to calculate the effective refractive index of the SPP mode in air, $n_{eff}^{air}$, for a given device from the Bragg condition for a quarter-wave stack, $2(d_{air} n_{eff}^{air} + d_{PMMA} n_{eff}^{PMMA}) = \lambda_0$. Here $d_{air}$ and $d_{PMMA}$ are the widths of the slabs in air and PMMA, respectively. Figure S1(a) shows the calculated $n_{eff}^{air}$ for devices of varying nanowire diameter compared to simulated values from FDTD. The measured values are within about 2% of the calculated values across the range of nanowire diameters we measured.

Fig. S1(b) shows the dependence of stopband position on DBR period. Using the calculated diameter-dependent $n_{eff}^{air}$ for each device, we plot the stopband position $\lambda_0 / n_{eff}^{air}$ and find that it matches the Bragg condition for a quarter wave stack.

## C. Theoretical limits of $Q$

The quality factor ($Q$) of a resonator can be expressed in terms of the separate loss channels in the cavity:

$$\frac{1}{Q} = \frac{1}{Q_{SP}} + \frac{1}{Q_{ff}} + \frac{1}{Q_R} \tag{2}$$

where $Q_{SP}$ results from material absorption of the SPP mode, $Q_{ff}$ is due to far-field scattering, and $Q_R$ is due to sub-unity reflection at the mirrors. $Q_{SP}$ represent the theoretical limit of Q that is limited only by the material absorption. $Q_{SP}$ is defined as the energy stored divided by the power dissipated:

$$Q_{SP} = -\frac{\omega E}{dE/dt}, \tag{3}$$

which can be rearranged to give the following relationship:

$$E = E_0 \exp\left(-\frac{\omega t}{Q_{SP}}\right) \tag{4}$$

The lifetime of a plasmon in the cavity, $\tau_0$, is therefore given by

$$\tau_0 = \frac{Q_{SP}}{\omega} \tag{5}$$

Expressing $\tau_0$ in terms of the SPP propagation length on resonance ($L_{SPP}$) and the phase velocity ($v_p$) and substituting for $\omega$, we get

$$\frac{L_{SPP}}{v_p} = \frac{Q_{SP}}{2\pi v_p / \lambda_{SP}} \tag{6}$$

$$Q_{SP} = 2\pi \frac{L_{SPP}}{\lambda_{SP}} \tag{7}$$

For a 100 nm diameter silver nanowire, we have previously measured the propagation length at 650 nm to be around 5 μm, which would put an upper bound on $Q$ of 101. We find that our resonators have $Q$ approaching 100, indicating that for our best devices losses are dominated by propagation length in the material rather than imperfections in the mirrors or device geometry.

**D. Quantum dot lifetime on silver nanowires**

Time-resolved fluorescence measurements of quantum dots coupled to silver nanowires surrounded in unpatterned PMMA show a multi-exponential decay, with the initial slope corresponding to a lifetime of 4 ns (Fig. S2). This indicates that the nonradiative and radiative emission enhancement due to the silver nanowire contributes a factor of ~4 to the overall lifetime reduction.

**E. Quantum dot mixture fluorescence modification**

In order to investigate the radiative contribution to Purcell enhancement, we coupled a mixture of CdSe quantum dots spanning an emission range of 550-800 nm to the resonators by incorporating them into the PMMA before device fabrication. The transmission and fluorescence spectra are shown in Fig. S3. The plasmon-coupled fluorescence emission exhibits oscillations that match the transmission spectrum, and the intensity at the cavity resonance is three to four times higher than the intensity outside of the stopband, while the transmission intensity on resonance is about half the intensity outside of the stopband. This observation provides another signature of radiative Purcell enhancement, in addition to the direct lifetime modification measurement described in the main text.

**F. Fluorescence of NVs coupled to silver nanowires**

NVs coupled to silver nanowires can emit into the far field or into SPP modes. The far-field fluorescence spectrum is essentially unmodified (Fig. S4(a)), while the SPP-coupled fluorescence exhibits small oscillations originating from Fabry-Perot resonance in the silver nanowire (Fig. S4(b)). As described in the main text, SPP-coupled NV fluorescence is modified dramatically when the NV center is coupled to a plasmon cavity.

## G. Purcell enhancement and mode volume in plasmon resonators

The spontaneous emission rate of an emitter in a cavity, $\Gamma$, is given by Fermi's Golden Rule:

$$\Gamma = |\langle \mu \cdot E \rangle|^2 \int \rho_e(\omega)\rho_c(r,\omega) dr d\omega, \tag{8}$$

where $\mu$ is the transition dipole moment of the emitter, $E$ is the electric field, $\omega$ is the angular frequency, $r$ is the position of the emitter, $\rho_e$ is the optical density of states of the emission, and $\rho_c$ is the optical density of states of the environment. In the weak coupling limit, an emitter placed at the peak electric field of the cavity mode exhibits a Purcell enhancement of spontaneous emission, $F$, given by

$$F = \frac{\Gamma}{\Gamma_0} = \frac{3\pi^2 Q}{4 V_{eff}}. \tag{9}$$

Here $\Gamma$ is the spontaneous emission rate in the cavity, $\Gamma_0$ is the rate in a uniform dielectric medium, $Q$ is the quality factor of the cavity, and $V_{eff} = \left(\frac{n}{\lambda_0}\right)^3 \frac{\int \varepsilon |E|^2 dV}{\max(\varepsilon |E|^2)}$ is the dimensionless mode volume. In an absorbing, dispersive medium such as silver, the mode volume can be estimated using the Drude model result for the energy density [4-6].

We simulated a plasmon resonator made of a 100 nm diameter silver nanowire using FDTD (Lumerical): in particular, this resonator was designed for a $\lambda_{SPP}$-mode with a resonant wavelength of 630 nm. Evaluating the expression using the electric field intensity at the resonant wavelength from simulations, we calculate a dimensionless $V_{eff}$ of 0.04 $(\lambda_0/n)^3$. We can also define an effective mode area,

$A_{eff} = \int \varepsilon |E|^2 dA / \max(\varepsilon |E|^2)(n/\lambda_0)^2$, which is 0.032 $(\lambda_0/n)^2$ from the simulations. These values yield an effective cavity length, $L_{eff}$ of 1.1875 $(\lambda_0/n)$ = 1.57 $\lambda_{SPP}$.

The overall Purcell enhancement can be decomposed into two components. One is associated with sub-wavelength, transverse confinement of guided SPPs, while another is associated with longitudinal confinement due to plasmonic cavity. The Purcell enhancement for an emitter placed next to a silver nanowire without a cavity is determined by the effective area of the SPP mode ($A_{eff}$) and the density of states ($dk/d\omega$) [7]:

$$F_0 \propto \frac{1}{A_{eff}} \frac{dk}{d\omega} = \frac{1}{A_{eff}} \frac{1}{v_g} \qquad (10)$$

We note since that the finesse of the cavity ($\mathfrak{I}$) is given by $L_{SPP}/L_{eff}$, and $\lambda_{SPP} \sim v_g$, the Purcell enhancement in the resonator can be written in terms of the enhancement due to the bare wire: $F \propto F_0 \mathfrak{I}$. This observation indicates that the cavity introduces a factor of the finesse to the enhancement compared to the silver nanowire. This intuitive result shows that the mode volume confinement is due to the radial dimension of the wire, and the Purcell factor can be further increased by the number of roundtrips of the plasmon in the cavity.

To understand the scaling of Purcell enhancement with nanowire dimensions, we note that in the quasistatic limit, $A_{eff}$ scales as $R^2$, and the group velocity scales as $R$, giving an overall scaling of $1/R^3$. For plasmonic resonators, $V_{eff} \propto A_{eff} L_{eff}$ and $Q = 2\pi \frac{L_{SPP}}{\lambda_{SP}} \cdot L_{eff}$ can be chosen to be as small as $\lambda_{SPP}/2$. Since $\lambda_{SPP}$ and $L_{SPP}$ both scale as $R$, the overall scaling of the Purcell enhancement is still $1/R^3$ for the resonator case. The Purcell enhancement can thus be significantly higher for thinner nanowires.

**H. Purcell enhancement of a broadband emitter**

When an emitter with emission rate $\Gamma_0$ and bandwidth $\Delta\omega$ such that $\Delta\omega > \frac{1}{\Gamma_0}$ is coupled to a cavity mode with decay rate $\kappa = \omega/Q < \Delta\omega$, the total emission rate is given by

$$\Gamma = \Gamma_{freespace} + \Gamma_{cavity} = \Gamma_0 + \Gamma_0 \frac{g^2}{\kappa \Gamma_0} \frac{\kappa}{\Delta\omega} \tag{11}$$

where $g = \sqrt{\frac{\omega \mu^2}{2\hbar\varepsilon V_{eff}}}$ is the coupling strength between the emitter and the cavity. For emission into the cavity mode to be greater than emission into free space, we have the condition

$$\frac{g^2}{\Gamma_0 \Delta\omega} > 1 \tag{12}$$

This condition is independent of $\kappa$, and therefore independent of $Q$. An enhancement of the emission rate therefore requires that $g$ be increased, which can be achieved by decreasing $V_{eff}$. We note that this is true only for emitters that have an incoherent contribution to emission and therefore do not have transform-limited linewidths ($\Delta\omega > \frac{1}{\Gamma_0}$).

**I. Determination of radiative Purcell enhancement of single emitter**

In order to quantify the contribution of the cavity to the Purcell enhancement of the NV center, we compared the intensity of the SPP-coupled emission on resonance with the intensity outside of the stopband, as described in the main text. This approximation is complicated by several other factors, but we chose a range of comparison that gave us the most conservative estimate of the Purcell enhancement. First, sideband oscillations evident in Figs. 1 and 2 show can also lead to an enhancement of the decay. Our calculated Purcell factor does not take this enhancement into account, so the actual contribution of the cavity could be higher. Second, we account for the difference in transmission and out-coupling losses at the different wavelengths by calibrating with the transmission spectrum. From equation (1) above, the ratio of signal in the red and the blue from the transmission spectrum is given by $\frac{I_{red}}{I_{blue}} \propto e^{-l/l_0(red)-l_0(blue)}$. The NV emission experiences a shorter path length, thus decreasing this ratio. Therefore, our calculated Purcell factor underestimates the enhancement by a factor of $e^{-l_{NV}/l_{wire}}$ where $l_{wire}$ is the length of the wire, and $l_{NV}$ is the path length from the NV to the end of the wire.

FIG. S1. (a) Simulated (blue) and measured (red) $n_{eff}^{air}$ as a function of diameter for various devices. The simulations and measurements agree to within 2%. (b) Stopband position ($\lambda_0$) versus DBR period for various devices. The silver NW diameter varies among devices, so the stopband position is corrected by the simulated value for $n_{eff}^{air}$.

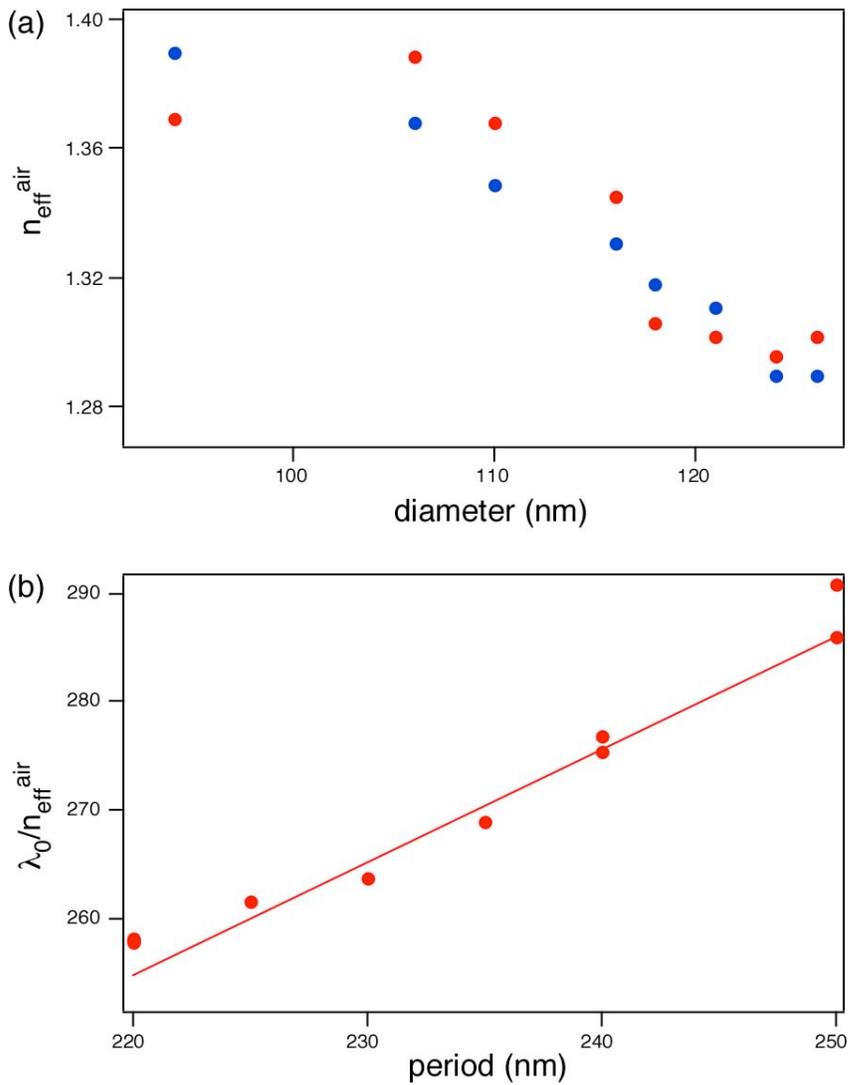

FIG. S2. Lifetime measurements of CdSe quantum dots on a silver NW surrounded by unpatterned PMMA.

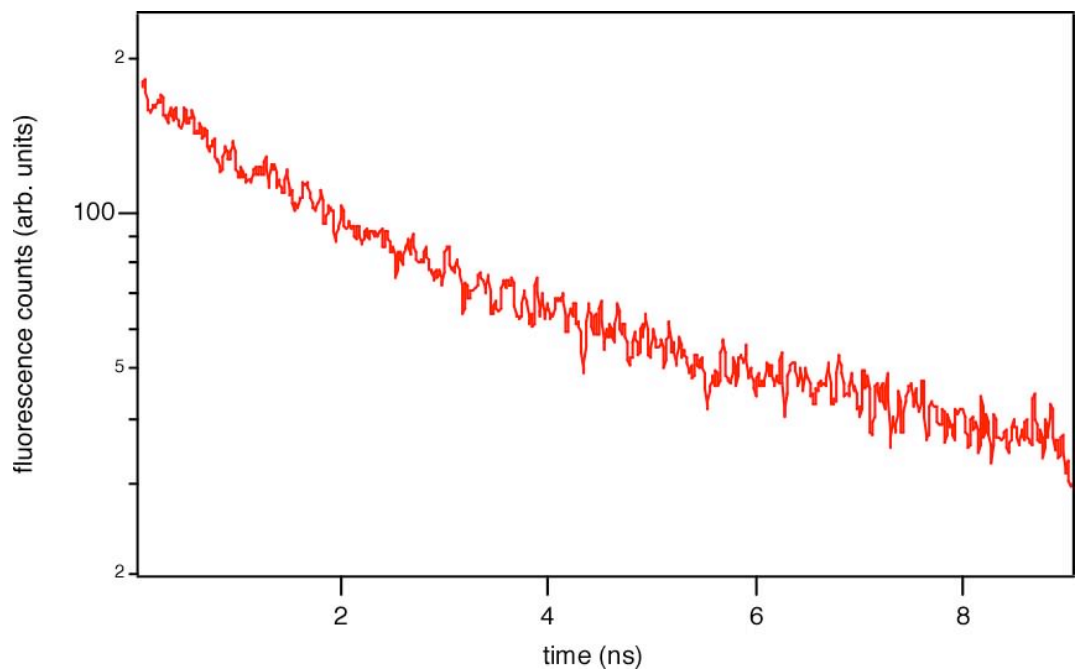

FIG. S3. Fluorescence intensity (red) collected from the end of the wire during excitation of the other end in green, and transmission spectrum (black). The oscillations in fluorescence match the sideband oscillations in transmission, as well as the resonance peak. The fluorescence intensity is higher on resonance than outside of the stopband, indicating Purcell enhancement of spontaneous emission.

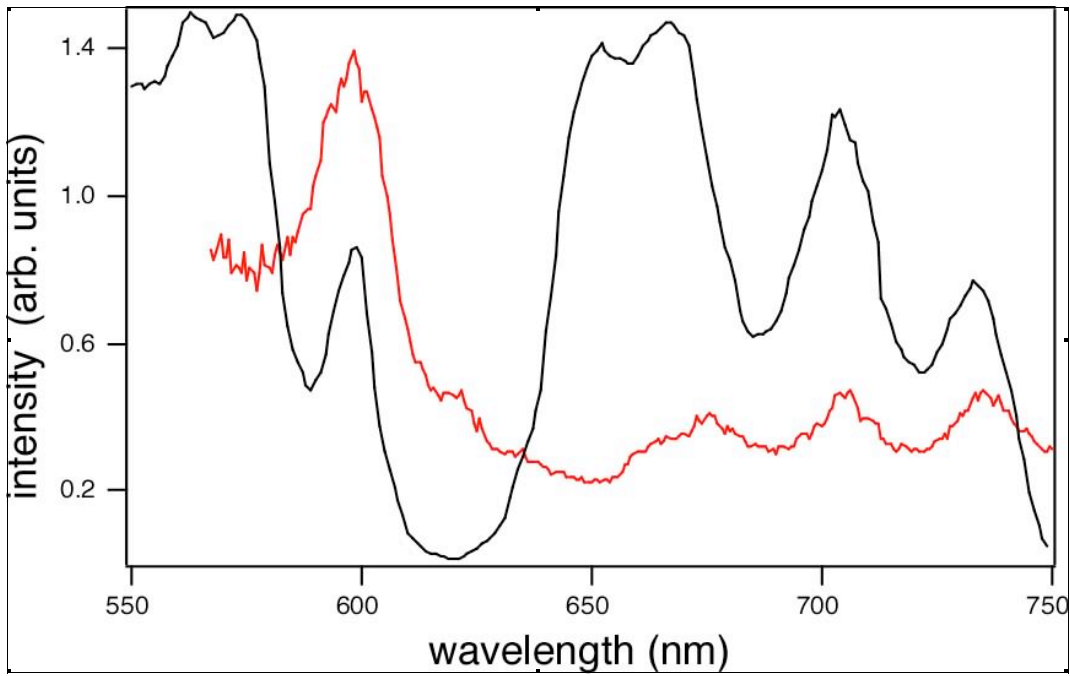

FIG. S4. Far-field (a) and SPP-coupled (b) fluorescence spectra of an NV coupled to a silver NW. Far-field emission is broad and essentially unmodified compared to the uncoupled NV, while SPP-coupled fluorescence exhibits small oscillations due to Fabry-Perot resonance in the silver NW. (c) SEM images of diamond nanocrystals coupled to a silver NW before (left) and after (right) resonator fabrication. Scale bars = 1μm.

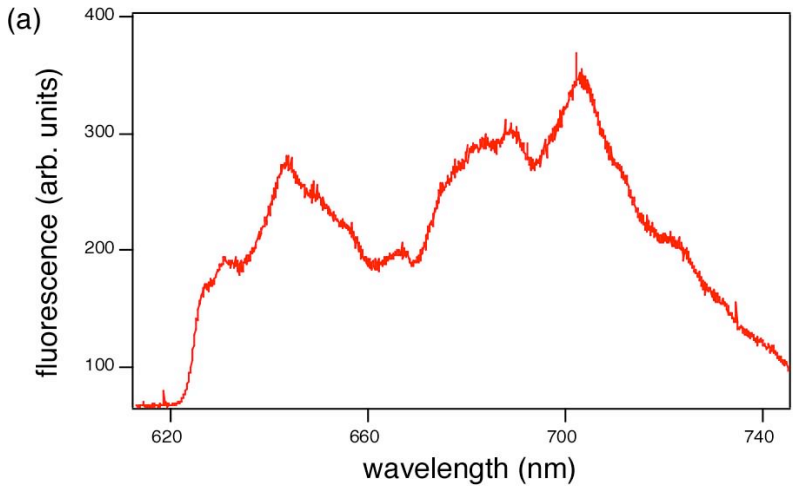
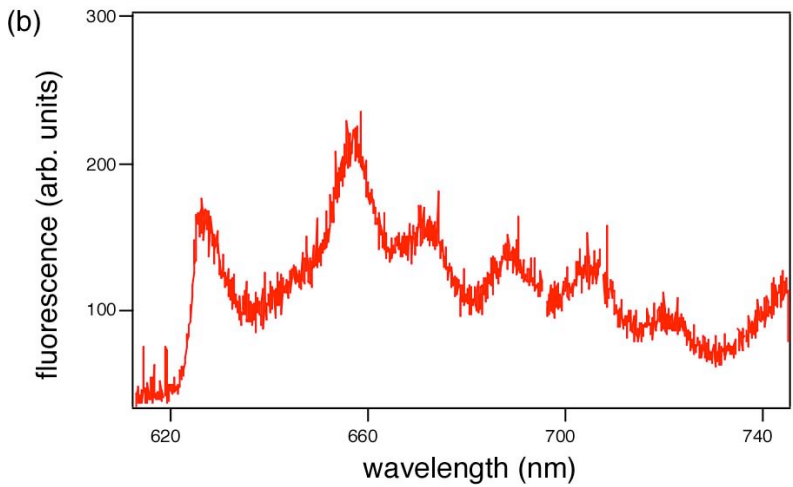
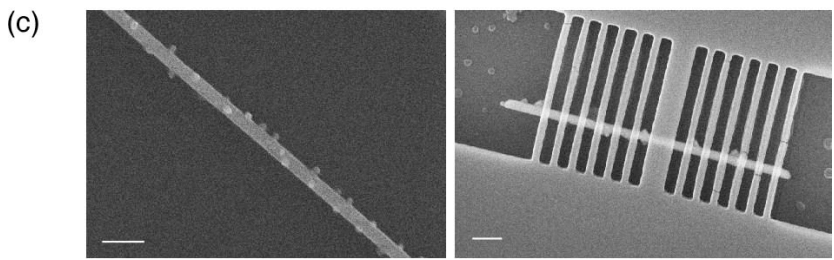